\documentstyle[prl,aps,multicol,epsf]{revtex}

\begin{document}
\input epsf 
\title{
Polarization of Bloch electrons and Berry phase in 
the presence of electromagnetic fields}

\author{Jun Goryo and Mahito Kohmoto} 
\address{\it Institute for Solid State Physics, University of Tokyo,  
5-1-5 Kashiwanoha, Kashiwa, Chiba, 277-8581 Japan}

\date{\today}
\maketitle

\begin{abstract}

We consider Bloch electrons in the 
presence of the uniform electromagnetic field in two- and three-dimensions. 
It is renowned that the quantized Hall effect occurs in 
such systems. We suppose a weak and homogeneous electric field 
represented by the time-dependent vector potential which 
is changing adiabatically. The adiabatic 
process can be closed in the parameter space and a Berry phase is generated.  
In the system, one can define the macroscopic electric polarization whose time 
derivative is equivalent to the quantized Hall current and its
 conductivity is written by the Chern number.  
Then, the polarization is induced perpendicular 
to the electric field. We show that the induced polarization per a cycle in the
 parameter space is quantized and closely related to the Berry phase as well as 
the Chern number. The process is adiabatic and the system always remains
 the ground state, then, 
the polarization is quite different from the usual dielectric polarization 
and has some similarity to the spontaneous polarization in the crystalline dielectrics which is 
also written by the Berry phase.     
We also point out the relation between our results and 
the adiabatic pumping. 

\end{abstract}

\begin{multicols}{2}

\section{Introduction}

The presence of the geometrical phase (Berry phase) was revealed 
in the adiabatic process of a quantum mechanical system around 
a closed loop in a parameter space\cite{Berry}. Thus, it could be
 regarded as the generalization of the Aharonov-Bohm effect\cite{A-B} in 
a parameter space. 
The Berry phase has been appeared in many contexts, for example, 
a treatment of the Born-Oppenheimer approximation\cite{BOA}, 
fractional statistics\cite{fractional-statistics}, the axial anomaly in 
field theories\cite{anomaly}, and so on.  

Recently, the relation between the spontaneous electric polarization 
in crystalline dielectrics and the Berry phase was discussed
\cite{King-Smith-and-Vanderbilt-Resta}. 
The authors of Ref.\cite{King-Smith-and-Vanderbilt-Resta} 
treated the system as Bloch electrons with a finite energy gap {\it in
the absence of} electromagnetic 
fields.  They introduce an adiabatic change of the 
potential (Kohn-Sham potential) with a slowly varying 
parameter $\lambda$. 
The Hamiltonian is compactified in the parameter space,  
and then, a Berry phase is defined in the adiabatic process. 
The charge transfer occurs and the polarization is induced 
in this process with external electric field 
held to be zero, and it is represented by 
the Berry phase\cite{King-Smith-and-Vanderbilt-Resta}.  

The two-dimensional (2D)  
Bloch electron system with an uniform magnetic field 
has been investigated for a long time\cite{2D-Bloch-in-B},
and it has been revealed that the integer quantum Hall effect  
occurs when the Fermi energy lies in an energy gap\cite{TKNN}.   
The integer part of the quantized Hall 
conductance $n e^2 / h$ ($n=0, \pm1, \pm2 
\cdot\cdot\cdot$) is represented by the Chern number\cite{Kohmoto-85,topology-text}, 
which is a topological number defined on the two-torus (the 
magnetic Brillouin zone). The effect is generalized 
to 3D\cite{Kohmoto-Halperin-Wu,3D-QHE}. Recent arguments point out 
that the band gap could exist for the magnetic field around 40 Tesla in 
organic compounds (TMTSF)$_2$X \cite{3D-QHE,organic}. Then, to realize the quantum  
Hall effect of Bloch electrons in 3D may be easier than that in 2D.

In this paper, we consider Bloch electrons in the presence of the  
uniform electromagnetic fields in 2D and 3D. 
A Berry phase is induced by the adiabatic change of 
the time-dependent vector potential.  
Following Ref.\cite{King-Smith-and-Vanderbilt-Resta}, 
we define the electric macroscopic polarization in the system.  
It is shown that the time derivative of the macroscopic electric 
polarization corresponds to the quantized Hall current in the system.
The quantized Hall current and also the polarization is 
induced adiabatically. In the adiabatic process, the system 
always remains to be in the ground state, and then, the polarization 
is quite different from the usual dielectric polarization. 
It has been pointed out that the quantized Hall conductivity 
is written in terms of the Berry phase as well as the Chern number in the 2D systems\cite{Kohmoto-93}. 
Recently, the authors generalize the relation  
to 3D systems\cite{Goryo-Kohmoto-1}. 
By using the relation between the Berry phase and 
the quantized Hall conductance, 
we can find out that the macroscopic polarization is closely
related to the Berry phase. The relation between the macroscopic polarization 
and the Berry phase seems to be analogous to the spontaneous polarization in the crystalline 
dielectrics\cite{King-Smith-and-Vanderbilt-Resta}. 

Recently, the adiabatic pumping is discussed 
progressively\cite{Thouless-pump,Thouless-pump-2}. 
In pumping, an adiabatic ac perturbation yields a dc current, 
and the charge transfer per the cycle is independent on the period 
of the perturbation. We also argue the relation between our results 
and the adiabatic pumping. 
   
In section II, we consider the 2D system.  
3D system is discussed in section III.  
We set the light velocity $c=1$ throughout the discussion. 

\section{macroscopic polarization and Berry phase in 2D system}

In this section, we treat 2D noninteracting electrons in a 
periodic potential in the presence of a uniform magnetic field 
perpendicular to the plane and a uniform electric field 
in the plane. The electric field 
is represented by a time-dependent vector potential. Then we employ 
the adiabatic approximation assuming that the electric field is weak
enough. In order to make the paper self-contained and fix the notations,
we would like to introduce the discussions in Ref.\cite{Kohmoto-93},
first. We review that the existence   
of the Berry phase is demonstrated in this 
situation and that the Chern number is written in terms of the Berry
phase\cite{Kohmoto-93}. The Chern number is well known as a topological number 
which represents the integer part of the quantized Hall conductance 
in the system\cite{TKNN,Kohmoto-85,topology-text}. 
After these reviews, we
show that the macroscopic electric polarization, whose 
time derivative corresponds to the Hall current in the system, 
is written in terms of the Berry phase. 

The time-dependent Shr{\"o}dinger equation we consider here is 
\begin{equation}
i \hbar \frac{\partial \Psi (t)}{\partial t}  = H(t) \Psi (t), 
\label{t-dep-sh-eq}
\end{equation}
where 
$$
H(t)=\frac{1}{2m} (- i \hbar {\bf \nabla} + e {\bf A}(t))^2 + U({\bf r}), 
$$
and ${\bf A}(t)$ is the vector potential for the electromagnetic field, 
therefore, the magnetic field and 
the electric field are given by ${\bf B} = {\bf \nabla} \times {\bf A}$ and 
${\bf E} = - \partial {\bf A} / \partial t$, respectively. 
We take the Bravais lattice vectors as 
\begin{equation}
{\bf R}=m {\bf a} + n {\bf b}, 
\end{equation}
where $m$ and $n$ are integers,   
and assume that the potential 
is periodic in both ${\bf a}$- and ${\bf b}$- directions, i.e. 
\begin{equation}
U({\bf r})=U({\bf r} + {\bf a})=U({\bf r} + {\bf b}). 
\end{equation}
We suppose that the electric field is weak enough. Then the system
evolves slowly so that it can be described by the adiabatic
approximation. At any instant, we have an eigenvalue equation 
\begin{eqnarray}
H(t) \Phi_n(t, {\bf r})
&=&\left[\frac{1}{2 m} (- i \hbar {\bf \nabla} + e {\bf A}(t))^2 + 
U({\bf r})\right] \Phi_n(t, {\bf r})
\nonumber\\
&=&E_{n} (t) \Phi_n(t, {\bf r}). 
\label{instant-shoroedinger}
\end{eqnarray}
The system is invariant under translations 
by ${\bf a}$ and by ${\bf b}$. 
However, the Hamiltonian $H(t)$ is not invariant 
under these transformations. The reason for this is that the vector 
potential ${\bf A}(t)$ is not constant in space in spite of the fact that 
the magnetic field is uniform. An appropriate gauge transformation is 
required to make the Hamiltonian invariant. It is well known that we can 
construct the magnetic translation operators, which commute with
Hamiltonian, written as    
\begin{equation}
T_{\bf R}=\exp\left[i {\bf R} \cdot \left(- i {\bf \nabla} 
- \frac{e}{\hbar} {\bf A}(t)\right)\right].    
\label{MBT}
\end{equation}
Here we use the symmetric gauge ${\bf A}=({\bf B} \times {\bf r}) / 2$. 
Now we look for eigenstates which simultaneously diagonalize $T_{\bf R}$ 
and $H$. However, note that the magnetic translations do not commute
with each other in general, since 
\begin{equation}
T_{\bf a} T_{\bf b} = \exp[2 \pi i \phi]T_{\bf b} T_{\bf a},   
\end{equation}
where $\phi=(e B / h) |{\bf a}\times {\bf b}|$ is the number of magnetic flux quanta 
in a unit cell. When $\phi$ is a rational number $\phi=p / q$ ($p$ and
$q$ are integers with no common factor), we have a subset of
translations which commute with each other. We consider the enlarged
unit cell (the magnetic unit cell), in which integral magnetic flux
quanta go through. For example, if we take the Bravais lattice vectors
as   
\begin{equation}
{\bf R}^{\prime}=m {\bf a} + n q {\bf b}, 
\end{equation}
then $p$ of the magnetic flux quanta penetrates the magnetic unit cell 
which is enclosed by the vectors ${\bf a}$ and $q {\bf b}$. The
magnetic translation operators $T_{\bf R^{\prime}}$ which corresponds 
to the new Bravais lattice vectors commute with each other. 

Let $\Phi^{(\alpha)}$ be an simultaneous eigenfunction of the operators 
$H$ and $T_{{\bf R}^{\prime}}$, here $\alpha$ denotes the band index. 
Then the eigenvalues of $T_{\bf a}$ and
$T_{q {\bf b}}$ are given by 
\begin{eqnarray}
T_{\bf a} \Phi^{(\alpha)} &=& e^{i {\bf k} \cdot {\bf a} } \Phi^{(\alpha)}, 
\nonumber\\
T_{q {\bf b}} \Phi^{(\alpha)} &=& e^{i {\bf k} \cdot q {\bf b} } 
\Phi^{(\alpha)}, 
\label{MBTeigenvalue}
\end{eqnarray}
where $\hbar {\bf k}$ is the generalized crystal momentum 
in the magnetic Brillouin zone (MBZ). 
By using the primitive vectors of the reciprocal lattice 
\begin{eqnarray}
{\bf G}_a&=&\frac{2 \pi}{v_0}({\bf b} \times \hat{\bf z}), 
\nonumber\\
{\bf G}_b&=&\frac{2 \pi}{v_0}(\hat{\bf z} \times {\bf a}),
\nonumber\\
v_0&=&\hat{\bf z}\cdot({\bf a}\times{\bf b}), 
\end{eqnarray}
where $\hat{\bf z}$ is the unit vector perpendicular to the 2D plane,
${\bf k}$  is written as 
\begin{eqnarray}
{\bf k}&=&f_a {\bf G}_a + \frac{f_b}{q} {\bf G}_b 
\nonumber\\
&&0 \leq f_a \leq 1, ~0 \leq f_b \leq 1.
\label{MBZ-2D}
\end{eqnarray}
Therefore, the eigenfunction is labeled by ${\bf k}$ in addition to
the band index $\alpha$.  
For the magnetic translation invariance, we can write
down the eigenfunction in the Bloch form 
\begin{eqnarray}
\Phi^{(\alpha)}_{\bf k}({\bf r}, t)=e^{i {\bf k}\cdot{\bf r}} u^{(\alpha)}_{\bf k} ({\bf r}, t).  
\label{Bloch}
\end{eqnarray}
From Eq.(\ref{MBT}) and Eq.(\ref{MBTeigenvalue}),  
$u^{(\alpha)}_{\bf k} ({\bf r}, t)$ should satisfy the relation 
\begin{eqnarray}
u^{(\alpha)}_{\bf k} ({\bf r} + {\bf a}, t)&=&
\exp\left[i \frac{e}{\hbar}{\bf a}\cdot{\bf A}(t) \right] 
u^{(\alpha)}_{\bf k} ({\bf r}, t), 
\nonumber\\
u^{(\alpha)}_{\bf k} ({\bf r} + q {\bf b}, t)&=&
\exp\left[i \frac{e}{\hbar} q{\bf b}\cdot{\bf A}(t)\right] 
u^{(\alpha)}_{\bf k} ({\bf r}, t).  
\label{Bloch-periodic}
\end{eqnarray}
This is the generalized Bloch theorem in the magnetic field\cite{2D-Bloch-in-B}. 
The instantaneous eigenvalue equation for $u^{(\alpha)}_{\bf k}({\bf r}, t)$,
which is obtained from Eq. (\ref{instant-shoroedinger}) and
Eq. (\ref{Bloch}) is written as 
\begin{eqnarray}
H_{\bf k}(t) u^{(\alpha)}_{\bf k}({\bf r}, t)&=&
E^{(\alpha)}_{\bf k}(t) u^{(\alpha)}_{\bf k}({\bf r}, t), 
\label{instant-shoroedinger-for-u}\\
H_{\bf k}(t)&=&\frac{(- i \hbar \nabla + \hbar {\bf k} + 
e {\bf A}(t))^2}{2 m} + U({\bf r}). 
\label{instant-hamiltonian}
\end{eqnarray}
Now, we have the ``Hamiltonian'' $H_{\bf k}(t)$, which contains the
parameters ${\bf k}$ and $t$. Note that the eigenvalue $E^{(\alpha)}_{\bf k}(t)$
depends on ${\bf k}$ continuously. For a fixed band index $\alpha$, the set of
values of $E^{(\alpha)}_{\bf k}(t)$ with ${\bf k}$ varying in the MBZ
forms the magnetic subband. 

Following Ref. \cite{Berry}, let us write a time-dependent 
wave function in the adiabatic approximation 
in terms of a snap shot wave function with the dynamical phase $\int_0^t
d t^{\prime} E_{\bf k}(t^{\prime})$ and a phase $\gamma^{(\alpha)}_{\bf k}(t)$  
\begin{eqnarray}
\Psi^{(\alpha)}_{\bf k}(t)=\exp\left[- \frac{i}{\hbar} 
\int_0^t E^{\alpha}_{\bf k} (t^{\prime}) dt^{\prime} \right] 
\nonumber\\
\times \exp\left[\frac{i}{\hbar} \gamma^{(\alpha)}_{\bf k} (t)\right] \Phi^{(\alpha)}_{\bf k} (t).  
\label{snap-shot}
\end{eqnarray}
Substitute the above form into the time-dependent Shr{\" o}dinger
equation Eq.(\ref{t-dep-sh-eq}), then, the phase $\gamma^{\alpha}_{\bf k}(t)$ is written as, 
\begin{eqnarray}
\gamma_{\bf k}(t)&=&i 
\int_0^t \left< \Phi_{\bf k} (t^{\prime})\right| \frac{d}{d t^{\prime}} \left| 
\Phi_{\bf k} (t^{\prime}) \right> d t^{\prime}  
\nonumber\\
&=& i \int_0^t 
\left< u_{\bf k} (t^{\prime})\right| \frac{d}{d t^{\prime}} \left| 
u_{\bf k} (t^{\prime}) \right> d t^{\prime},  
\label{gamma2d}
\end{eqnarray}
here we omit the band index $\alpha$ because we will consider 
a single magnetic subband for a while. 
The ``wave function'' $u_{\bf k}$ 
obeys the ``Schr{\"o}dinger equation'' (\ref{instant-shoroedinger-for-u}). 

In order to consider a Berry phase, a Hamiltonian must go around a
closed loop in the parameter space in the adiabatic process. 
The Hamiltonian  Eq. (\ref{instant-hamiltonian}) as it is does not have
this property. However, it is possible to compactify it as 
\begin{equation}
H_{{\bf k}}(t) \sim H_{{\bf k} + {\bf G}_a} (t) 
\sim H_{{\bf k} + {\bf G}_b / q}(t), 
\end{equation}
where the symbol $\sim$ denotes equivalence, 
since we have equivalent crystal momenta and thus give the same wave
function. The time-dependent part of the vector potential is written as 
$- {\bf E} t$. Suppose that the electric field is applied along 
${\bf G}_a \perp {\bf b}$. Then the time-dependence of the Hamiltonian 
$H_{\bf k}(t)$ enters in the form ${\bf k} - (e E v_0 t / h b) {\bf G}_a$. Thus we have 
$u_{\bf k}(t)=u_{{\bf k} - (e E v_0 t / h b) {\bf G}_a}$, where $u_{\bf
k}(t=0)=u_{\bf k}$, and the period of a transport around a closed
path in the parameter space is given by $T=(h b/ e E v_0)$\cite{Zak}. 
The Berry phase is obtained from Eq. (\ref{gamma2d}) as 
\begin{eqnarray} 
\Gamma_a(f_b)&=&i \int_0^T \left< u_{{\bf k} - t^{\prime}\frac{e E v_0}{h b}{\bf G}_a} \right| 
\frac{\partial}{\partial t^{\prime}} \left| u_{{\bf k} - t^{\prime}\frac{e E v_0 }{h b}{\bf G}_a} 
\right> d t^{\prime}
\nonumber\\
&=&i \int_0^1 \left< u_{\bf k} \right| 
\frac{\partial}{\partial f_a} \left| u_{\bf k} \right> d f_a,  
\label{Gamma-y}
\end{eqnarray}
where $f_a$ and $f_b$ are defined in Eq. (\ref{MBZ-2D}). 
$\Gamma_a(f_b)$ depends on $f_b$ but is independent on $f_a$. 
Similarly, if the electric field is applied along ${\bf G}_b \perp {\bf a}$, 
we have 
\begin{eqnarray}
\Gamma_b(f_a)&=&i \int_0^1 \left< u_{\bf k} \right| 
\frac{\partial}{\partial f_b} \left| u_{\bf k} \right> d f_b .
\end{eqnarray}
In general, we can define $T$ and Berry phase 
when the electric field is parallel to the reciprocal lattice vector for
the magnetic unit cell,
i.e., 
\begin{equation}
{\bf E} ~//~ (m {\bf G}_a + n {\bf G}_b/ q),  
\label{commensulate}
\end{equation}
where $m,n=0,\pm1,\pm2\cdot\cdot\cdot$. 

Let us define a vector field in the MBZ by 
\begin{eqnarray}  
\tilde{\bf A}({\bf k})&=&\left<u_{\bf k} | {\bf \nabla}_{\bf k} | u_{\bf k} \right>,   
\label{Berry-connection}
\end{eqnarray}
where ${\bf \nabla}_{\bf k}=\left[{\bf a} \frac{\partial}{\partial f_a} + q
{\bf b} \frac{\partial}{\partial f_b} \right]$ is the vector operator. 
Then, we have 
\begin{equation}
\Gamma_a (f_b) = i \int_0^1 d f_a {\bf G}_a \cdot \tilde{\bf A}({\bf k}) = 
i \oint_{C(f_b)}   d {\bf k} \cdot \tilde{\bf A} ({\bf k}),   
\label{Berry-phase-x}
\end{equation}
where $\oint_{C(f_b)}$ denotes that the path of the line integral is taken on the closed loop 
where $f_b$ is fixed. Similarly we have 
\begin{equation}
\Gamma_b (f_a) = i \int_0^1 d f_b \frac{{\bf G}_b}{q} \cdot \tilde{\bf A}({\bf k}) = 
i \oint_{C(f_a)} d {\bf k} \cdot \tilde{\bf A} ({\bf k}).  
\label{Berry-phase-y}
\end{equation}
The vector field $\tilde{\bf A}({\bf k})$ is called Berry
connection. This is considered as a gauge field
induced in the parameter space, or in the MBZ. Suppose 
$u_{\bf k}({\bf r})$ satisfies the Schr{\"o}dinger equation
Eq. (\ref{instant-shoroedinger-for-u}), then an phase transformed function 
\begin{equation}
u^{\prime}_{\bf k}({\bf r}) = u_{\bf k}({\bf r}) e^{i f({\bf k})}, 
\end{equation}
also satisfies Eq. (\ref{instant-shoroedinger-for-u}), 
where $f({\bf k})$ is the smooth function of ${\bf k}$
and is independent of ${\bf r}$. Under the transformation, 
any physical variables do not change, since this transformation 
is global in the coordinate space. From Eq. (\ref{Berry-connection}), 
$\tilde{\bf A}({\bf k})$ is transformed as 
\begin{equation}
\tilde{\bf A}^{\prime}({\bf k})=\tilde{\bf A}({\bf k}) + i {\bf \nabla}_{\bf k} f({\bf k}).  
\label{Berry-connection-gauge}
\end{equation}
Thus, we may regard $\tilde{\bf A}({\bf k})$ 
as a gauge field induced in the parameter space. 

Let us introduce an integral 
\begin{equation}
N_{\rm Ch}=\frac{1}{2 \pi i} \int_{\rm MBZ} d^2 k 
\left[{\bf \nabla_{\bf k} \times \tilde{\bf A}({\bf k})}\right]_z.   
\label{Chern-number}
\end{equation}
Here, $[\cdot\cdot\cdot]_z$ expresses the $z$-component of the vector.
The MBZ is topologically a torus $T^2$ rather than a rectangular in ${\bf
k}$-space, because the points ${\bf k}$, ${\bf k} + {\bf G}_a$ and
${\bf k} + {\bf G}_b / q$ should be identified as equivalent points. 
Therefore, integral coincides with the first Chern number of a principal $U(1)$ fiber bundle on
the torus $T^2$ whose connection is $\tilde{\bf A}({\bf k})$\cite{Kohmoto-85,topology-text}.   
Since there are no boundary on the torus, the integral
(\ref{Chern-number}) becomes finite if and only if $\tilde{\bf A}({\bf k})$ has non-trivial
topology on the MBZ. 

We apply the Stokes theorem to Eq. (\ref{Chern-number}) and derive the relation between Berry phases
Eqs. (\ref{Berry-phase-x}), (\ref{Berry-phase-y}) and the Chern number
Eq. (\ref{Chern-number}) written as\cite{Kohmoto-93} 
\begin{equation}
N_{\rm Ch}=\frac{1}{2 \pi}\left[\int_0^1 d f_a 
\frac{d \Gamma_b(f_a)}{d f_a}  - 
\int_0^{1} d f_b \frac{d \Gamma_a (f_b)}{d f_b}  \right]. 
\label{Chern-Berry-2D}
\end{equation}



Following Ref.\cite{King-Smith-and-Vanderbilt-Resta}, 
one could define the electric contribution to 
the macroscopic electric polarization as, 
\begin{eqnarray}  
{\bf P}(t)&=&e \sum_{\alpha \leq \alpha_{\rm F}} 
\int_{\rm MBZ} \frac{d^2k}{(2 \pi)^2} 
\left< \Psi^{(\alpha)}_{\bf k} (t) 
\right| {\bf r} \left| 
\Psi^{(\alpha)}_{\bf k} (t) \right>
\nonumber\\
&=& e \sum_{\alpha \leq \alpha_{\rm F}} 
\int_{\rm MBZ} \frac{d^2k}{(2 \pi)^2} 
\left< u^{(\alpha)}_{\bf k} (t) 
\right| {\bf r} \left| 
u^{(\alpha)}_{\bf k} (t) \right>,   
\label{polar}
\end{eqnarray}
where we use Eqs. (\ref{Bloch}) and (\ref{snap-shot}), and 
the summation is taken over the bands below the Fermi energy. 
We reinstalled the band index $\alpha$. 
The time derivative of ${\bf P} (t)$ is the electric current of the
system. It should be the quantized Hall current when the Fermi energy
lies in the gap. Actually, it is calculated as,  
\begin{eqnarray}
\frac{\partial {\bf P}(t)}{\partial t}&=&e \sum_{\alpha \leq \alpha_{\rm F}} \int_{\rm MBZ} \frac{d^2k}{(2 \pi)^2} 
\left[
\left< \frac{\partial}{\partial t} u^{(\alpha)}_{\bf k} (t) \right| {\bf r} 
\left| u^{(\alpha)}_{\bf k} (t) \right> \right.
\nonumber\\
&&+
\left.\left<  u^{(\alpha)}_{\bf k} (t) \right| {\bf r} 
\left| \frac{\partial}{\partial t} u^{(\alpha)}_{\bf k} (t) \right>
\right]
\nonumber\\
&=&e \sum_{\alpha \leq \alpha_{\rm F}<\beta} \int_{\rm MBZ} \frac{d^2k}{(2 \pi)^2} 
\times
\nonumber\\
&&\left[
\left< \frac{\partial}{\partial t} u^{(\alpha)}_{\bf k} (t) \right| \left. u^{(\beta)}_{\bf k} (t)\right> 
\left< u^{(\beta)}_{\bf k} (t) \right|{\bf r}\left| u^{(\alpha)}_{\bf k} (t) \right> \right.
\nonumber\\
&& \left. +
\left<  u^{(\alpha)}_{\bf k} (t) \right| {\bf r} 
\left|u^{(\beta)}_{\bf k} (t)\right> 
\left< u^{(\beta)}_{\bf k} (t) \right. \left| \frac{\partial}{\partial t} 
u^{(\alpha)}_{\bf k} (t) \right>
\right], 
\nonumber\\
\label{dp/dt-1}
\end{eqnarray}
\noindent
where we use the completeness condition for $|u_{\bf k}^{\beta}(t)>$ 
$$
\sum_{\beta} \left|u^{(\beta)}_{\bf k} (t) \right> 
\left< u^{(\beta)}_{\bf k} (t) \right|=1,  
$$
and here, the summation is taken over all the bands. 
We can calculate the matrix element for $\alpha\neq\beta$ 
\begin{eqnarray}
&&\left< u^{(\beta)}_{\bf k} (t) 
\right|{\bf r}\left| u^{(\alpha)}_{\bf k} (t) \right>
\nonumber\\
&=&\frac{\left< u^{(\beta)}_{\bf k} (t) 
\right|[H(t), {\bf r}]
\left| u^{(\alpha)}_{\bf k} (t) \right>} 
{E^{(\beta)}_{\bf k} - E^{(\alpha)}_{\bf k}} 
\nonumber\\
&=&- i \hbar  
\frac{
\left< u^{(\beta)}_{{\bf k}} (t) \right| {\bf v}(t) \left| u^{(\alpha)}_{\bf k} (t) \right>}
{E^{(\beta)}_{{\bf k}} - E^{(\alpha)}_{{\bf k}}} 
\nonumber\\
&=&- i 
\frac{\left< u^{(\beta)}_{{\bf k}} (t) 
\right| \partial H_{\bf k}(t) / \partial {\bf k} \left| 
u^{(\alpha)}_{\bf k} (t) \right>}
{E^{(\beta)}_{{\bf k}} - E^{(\alpha)}_{{\bf k}}}
\nonumber\\
&=&- i 
\left< \frac{\partial u^{(\beta)}_{{\bf k}} (t)}{\partial {\bf k}} \right| \left. u^{(\alpha)}_{\bf k} (t) 
\right>
\nonumber\\
&=&i 
\left< u^{(\beta)}_{\bf k} (t) \right| \left.  \frac{\partial u^{(\alpha)}_{{\bf k}} (t)}
{\partial {\bf k}}\right>,  
\end{eqnarray}
where ${\bf v}(t) = (- i \hbar {\bf \nabla} + e {\bf A}(t)) / m$ is the
velocity operator. 
Then, we can find that 
\begin{eqnarray}
\frac{\partial {\bf P}}{\partial t}&=&
i e \int_{\rm MBZ} \frac{d^2 k}{(2 \pi)^2} 
\sum_{\alpha \leq \alpha_{\rm F}} 
\left[
\left< \frac{u^{(\alpha)}_{\bf k} (t)}{\partial t} \right| \left.  
\frac{\partial u^{(\alpha)}_{{\bf k}} (t)}{\partial {\bf k}}\right>- 
\right.
\nonumber\\
&&\left.\left< \frac{u^{(\alpha)}_{\bf k} (t)}{\partial {\bf k}} \right| \left.  
\frac{\partial u^{(\alpha)}_{{\bf k}} (t)}{\partial t}\right> \right]. 
\end{eqnarray}
We have an relation 
$|u_{\bf k} (t)>=| u_{{\bf k} - e {\bf E} t / \hbar } >$, thus 
$$
\frac{\partial u_{\bf k} (t)}{\partial t} =- \frac{e {\bf E}}{\hbar} \cdot 
\frac{\partial u_{\bf k}(t)}{\partial {\bf k}}. 
$$ 
Therefore, the time derivative  
of the macroscopic polarization is written as 
\begin{eqnarray}
&&\frac{\partial {\bf P}(t)}{\partial t}
\nonumber\\
&=&
i \frac{e^2}{\hbar} \sum_{i=x,y} E_i \int_{\rm MBZ} \frac{d^2 k}{(2 \pi)^2} 
\sum_{\alpha \leq \alpha_{\rm F}} \times 
\nonumber\\
&&\left[\left< \frac{u^{(\alpha)}_{\bf k} (t)}{\partial k_i} \right| \left.  
\frac{\partial u^{(\alpha)}_{{\bf k}} (t)}{\partial {\bf k}}\right> - 
\left< \frac{u^{(\alpha)}_{\bf k} (t)}{\partial {\bf k}} \right| \left.  
\frac{\partial u^{(\alpha)}_{{\bf k}} (t)}{\partial k_i}\right> \right]
\nonumber\\
&=&\frac{e^2}{h} {\bf E}\times \hat{\bf z}  
\left\{\int_{\rm MBZ} \frac{d^2 k}{2 \pi i} [\nabla_{\bf k} \times \tilde{\bf A}({\bf k})]_z \right\}
\nonumber\\
&=&\frac{e^2}{h} \left\{\sum_{\alpha \leq \alpha_{\rm F}}N^{(\alpha)}_{\rm Ch}\right\} {\bf E}\times\hat{\bf z}, 
\label{dp/dt-2}
\end{eqnarray}
where $N^{(\alpha)}_{\rm Ch}$ denotes the
Chern number for the $\alpha$-th band. Eq. (\ref{dp/dt-2}) is just the 
quantized Hall current in the system when the Fermi energy is located 
between the energy gap. 

Then, the polarization induced by the electric field along ${\bf
G}_a \perp {\bf b}$ per the period $T=(h b / e E v_0)$ (See, the paragraph before
Eq. (\ref{Gamma-y})) is written as 
\begin{eqnarray}  
\Delta {\bf P}_b &=&\int_0^T d t \frac{\partial {\bf P}_b(t)}{\partial t}
\nonumber\\
&=& -{q \bf b}\frac{e }{q v_0}  \sum_{\alpha \leq \alpha_{\rm F}} 
N^{(\alpha)}_{\rm Ch}
\nonumber\\ 
&=&-{q \bf b}\frac{e}{q v_0} \sum_{\alpha \leq \alpha_{\rm F}} \times 
\nonumber\\
&&\frac{1}{2 \pi}\left[\int_0^1 d f_a 
\frac{d \Gamma^{(\alpha)}_b(f_a)}{d f_a} 
 - \right.
\left. \int_0^{1} d f_b 
\frac{d \Gamma^{(\alpha)}_a (f_b)}{d f_b}  \right], 
\end{eqnarray}
where $\Gamma^{(\alpha)}_{a,b}$ is the Berry phase for the $\alpha$-th
band and we use Eq. (\ref{Chern-Berry-2D}). The result shows that 
the electric dipole moment per the magnetic unit cell 
is quantized as an integer multiple of $- e q {\bf b}$. 

Similarly, when ${\bf E} // {\bf G}_{b} \perp {\bf a}$, the period 
$T=h a / e E q v_0$ and 
\begin{equation}
\Delta {\bf P}_a={\bf a} \frac{e}{q v_0} \sum_{\alpha\leq\alpha_{\rm F}} N^{(\alpha)}_{\rm Ch}.    
\end{equation}
It is also written by the Berry phase by using
Eq. (\ref{Chern-Berry-2D}) as well as the Chern number. 
The electric dipole moment per the magnetic unit cell 
is integer multiple of $e {\bf a}$.  

One could point out the specific properties of the 
polarization in our system as follows; 
\begin{enumerate}

\item introduced by the {\it adiabatic} process 
\label{a}

\item perpendicular to {\bf E}
\label{b}

\item written by the Berry phase 
\label{c}

\item independent of $T$, i.e. independent of $|{\bf E}|$ 
\label{d}

\item the induced dipole moment per the magnetic unit cell per the period $T$ is quantized  
\label{e}

\end{enumerate}

From \ref{a} and \ref{b}, one can see that the polarization in our system 
is quite different in the usual dielectric polarization in the
insulators. In the adiabatic process, the system always remains in the 
ground state. In insulators, the longitudinal conductivity and  
the translational conductivity is zero, and the charge transport 
would not occur 
adiabatically. Then, the finite polarization is obtained beyond the adiabatic 
approximation and the direction of the polarization is along 
the external electric field ${\bf E}$. 
In our system, we have a finite transverse conductivity because 
of the presence of the magnetic field. The transverse charge transport 
occurs even in the adiabatic process 
and the polarization is induced perpendicular to ${\bf E}$. 
  
The properties \ref{a} and \ref{c} suggests that the
polarization in our system has some similarity to 
the spontaneous macroscopic polarization in the crystalline dielectrics.  
It has been pointed out 
by King-Smith and Vanderbilt, and 
Resta\cite{King-Smith-and-Vanderbilt-Resta} that the spontaneous 
polarization is induced by an adiabatic change of the 
potential (Kohn-Sham potential) with a slowly varying 
parameter $\lambda$. The process is defined 
on the closed loop in the parameter space and the Berry phase is
generated, and the spontaneous polarization is written by the Berry phase. 
A related result is obtained in a different context\cite{Ryu-Hatsugai}.   

We see the relation between our discussion and the adiabatic charge
pumping originally argued by Thouless\cite{Thouless-pump,Thouless-pump-2}.  
The adiabatic pumping is the charge transport whose specific 
properties are: (i) the adiabatic ac 
perturbation yields the dc current, (ii) the charge transfer does not
depend on the period of the perturbation\cite{Thouless-pump-2}. 

In our system, the perturbation (the electric field) itself is not ac,
but the Hamiltonian $H_{\bf k}(t)$ is compactified
in the parameter space and the period $T$ is introduced, and the dc Hall
current Eq. (\ref{dp/dt-2}) flows. 

The statements \ref{d} and \ref{e} imply the fact that the charge transfer per $T$
across the boundary of the magnetic unit cell does not depend on $T$, 
which corresponds to the property (ii) of the adiabatic pumping, 
and is quantized. 
Actually, one can see it directly. For ${\bf E} \perp {\bf b}$, 
the Hall current $\partial {\bf P}_b / \partial t$ flows parallel to
${\bf b}$ and across the boundary $a$-axis. The normal vector 
of the boundary is ${\bf G}_a / |{\bf G}_a|$. Then, the charge transfer is 
\begin{eqnarray}
\Delta Q&=&\int_0^T dt \int_0^1 d f_a \frac{{\bf G}_a}{|{\bf G}_a|} \cdot 
\frac{\partial {\bf P}_b}{\partial t}
\nonumber\\
&=&\int_0^1 d f_a \frac{{\bf G}_a}{|{\bf G}_a|} \cdot \Delta {\bf P}_b 
\nonumber\\
&=&- e \sum_{\alpha \leq \alpha_{\rm F}} N^{(\alpha)}_{\rm Ch}. 
\label{charge-2D}
\end{eqnarray}
For ${\bf E} \perp {\bf a}$, the charge transfer is 
\begin{equation}
\Delta Q=e \sum_{\alpha \leq \alpha_{\rm F}} N^{(\alpha)}_{\rm Ch}. 
\end{equation} 
Similar results are obtained by Thouless for the 
commensurate periodic ac perturbation\cite{Thouless-pump}. 
The commensurability seems to correspond to the direction of ${\bf E}$ in our 
discussion (See. Eq. (\ref{commensulate})), which is essential 
to define $T$ and $\Delta Q$.


\noindent
\section{macroscopic polarization and Berry phase in 3D system}

The aim of this section is to argue the 3D generalization of the
discussions in the previous section. 
%
We define the macroscopic polarization. 
One can see that its time derivative is equivalent to the quantized Hall 
current 
in 3D. The Hall conductivity is represented by the Chern number and  
quantized when the Fermi energy lies on the band
gap\cite{Kohmoto-Halperin-Wu}. Recent arguments point out 
that such a gap could exist for the magnetic field around 40 Tesla in  
organic compounds (TMTSF)$_2$X\cite{3D-QHE,organic}. Then, it may be  
easier to realize the quantum Hall effect of Bloch electrons 
in 3D rather than 2D. Our discussion written below could 
be applicable to such compounds by using the tight-binding scheme. 

The relation between the Berry phase and the Chern number 
is also driven in the system\cite{Goryo-Kohmoto-1}. 
Then, we would find out the relation between 
the macroscopic polarization and the Berry 
phase in 3D system as well as 2D system. 
We also point out the relation between our results and the adiabatic
pumping\cite{Thouless-pump,Thouless-pump-2}.

Let us consider the 3D Bloch electrons in the presence of uniform 
electromagnetic fields. We have a periodic potential 
$U({\bf r})=U({\bf r} + l {\bf a} + m {\bf b} + n {\bf c})$ ($l,m,n$;
integers). The electric field we consider here is weak and described 
by an adiabatically changing vector potential. We use the 
adiabatic approximation and first, we consider the eigenstates of the 
Hamiltonian $H(t)$ at fixed $t$. 

We have a magnetic field written as, 
\begin{equation}
{\bf B}=B_a {\bf a} + B_b {\bf b} + B_c {\bf c},  
\end{equation} 
where ${\bf a}, {\bf b}$ and ${\bf c}$ are the primitive 
vectors for the Bravais lattice. Let ${\bf G}_a=(2 \pi / v_0) 
({\bf b} \times {\bf c})$, ${\bf G}_b =(2 \pi / v_0) 
({\bf c} \times {\bf a})$ and 
${\bf G}_c =(2 \pi / v_0) ({\bf a} \times {\bf b})$
stand for the fundamental reciprocal lattice vectors, where 
$v_0={\bf a}\cdot({\bf b}\times{\bf c})$. 
$B_a$, $B_b$ and $B_c$ are written as, 
\begin{eqnarray}
B_a=\frac{1}{2 \pi}{\bf B}\cdot{\bf G}_a=\frac{1}{v_0}\frac{h}{e}\phi_a,  
\nonumber\\
B_b=\frac{1}{2 \pi}{\bf B}\cdot{\bf G}_b=\frac{1}{v_0}\frac{h}{e}\phi_b,  
\nonumber\\
B_c=\frac{1}{2 \pi}{\bf B}\cdot{\bf G}_c=\frac{1}{v_0}\frac{h}{e}\phi_c,  
\end{eqnarray}
where $\phi_a$, $\phi_b$ and $\phi_c$ is a number of the unit flux
through the plane ${\bf b} \times {\bf c}$, ${\bf c} \times {\bf a}$ 
and ${\bf a} \times {\bf b}$, respectively. 
We introduce a ``rational magnetic fields'', where    
$\phi_a=p_a / q_a$, $\phi_b= p_b / q_b$ and 
$\phi_c= p_c / q_c$ are rational values. Here $p_i$, and $q_i$
($i=a,b,c$) are the integers without common factor. Let $q$ stands for 
the least multiple factor for the three integers $q_a$, $q_b$ and
$q_c$. We can write $\phi_a = N_a / q$, $\phi_b = N_b / q$ and 
$\phi_c = N_c / q$ ($N_i$; integer, $i=a,b,c$). Let $p$ stands for the largest common factor for three
integers $N_a$, $N_b$ and $N_c$. The magnetic field is written as 
\begin{eqnarray}
{\bf B}&=&\frac{1}{v_0}\frac{h}{e} \frac{p}{q} {\bf c}^{\prime},  
\nonumber\\
{\bf c}^{\prime}&=&(n_a {\bf a} + n_b {\bf b} +n_c {\bf c}). 
\end{eqnarray}
By definition, $n_a$, $n_b$ and $n_c$ are integers with no common factor. 
It means that there are no vectors on the Bravais lattice which is 
the submultiple of ${\bf c}^{\prime}$, and it was shown by
Ref.\cite{Kohmoto-Halperin-Wu} that 
we can find vectors ${\bf a}^{\prime}$ and ${\bf b}^{\prime}$  
such that ${\bf a}^{\prime}$, ${\bf b}^{\prime}$ and ${\bf c}^{\prime}$
are a new set of primitive vectors as follows. 
Let $r$ stands for the greatest common factor of $n_c$ and $n_a$. 
Therefore, there is no common factor of $r$ and $n_b$. Then 
we can choose four integers $s_a$, $s_b$, $s_c$ and $s_r$, which 
satisfy relations   
$$
s_a n_a + s_c n_c =r,~ s_b n_b + s_r r =1,   
$$
and we may choose 
\begin{eqnarray}
{\bf a}^{\prime}&=&s_c {\bf a} - s_a {\bf c}, 
\nonumber\\
{\bf b}^{\prime}&=&s_r {\bf b} - \frac{s_b n_c}{r}{\bf c} - \frac{s_b n_a}{r}{\bf a}. 
\nonumber
\end{eqnarray}


The Hamiltonian has the magnetic translation symmetry. 
Actually, we can choose the magnetic unit cell with 
the primitive vectors ${\bf a}^{\prime}$, $ q {\bf b}^{\prime}$ and  
${\bf c}^{\prime}$, and the magnetic translation 
operators for the Bravais lattice 
$$
{\bf R}^{\prime}=l {\bf a}^{\prime}+ m q {\bf b}^{\prime} + n {\bf
c}^{\prime}
 ~~(l,m,n; ~{\rm integer}), 
$$
is written by the 3D generalization of Eq. (\ref{MBT}) when we take the
symmetric gauge. 
Three operators $T_{\bf a^{\prime}}$, $T_{q {\bf b^{\prime}}}$ 
and $T_{\bf c^{\prime}}$ commute with the Hamiltonian at any instant, and also with 
each other. The wavevector in the magnetic Brillouin zone (MBZ) is
written as 
\begin{eqnarray}
{\bf k}&=& f_{a^{\prime}} {\bf G}_{a^{\prime}} + \frac{f_{b^{\prime}}}{q} {\bf G}_{b^{\prime}} 
+ f_{c^{\prime}} {\bf G}_{c^{\prime}},   
~0 < f_{a^{\prime}}, f_{b^{\prime}}, f_{c^{\prime}} < 1. 
\label{crystal-momentum}
\end{eqnarray}
Then, the eigenstates of the Hamiltonian at any instant is the
generalized Bloch states. 
The Bloch wave function with a band index $\alpha$ 
is written as 
$\Phi^{\alpha}_{{\bf k}}({\bf r},t)=e^{i {\bf k}\cdot {\bf r}} u^{(\alpha)}_{{\bf k}}({\bf r},t)$, 
where $u_{\bf k}^{\alpha}({\bf r}, t)$ satisfies 
\begin{eqnarray}
&&H_{\bf k}(t) u^{(\alpha)}_{\bf k}(t, {\bf r})
\nonumber\\
&=&\left[\frac{1}{2 m} (- i {\bf \nabla} + {\bf k} + e {\bf A}(t))^2 
+ U({\bf r})\right] 
u^{(\alpha)}_{\bf k}(t, {\bf r})
\nonumber\\
&=&E^{(\alpha)}_{\bf k} (t) u^{(\alpha)}_{\bf k}(t, {\bf r}), 
\nonumber\\
&&{\bf A}(t)=- {\bf E}t + \frac{1}{2}{\bf B} \times {\bf r}, 
\end{eqnarray}
which is the 3D generalization of
Eqs. (\ref{instant-shoroedinger-for-u}) and 
(\ref{instant-hamiltonian}). 
The function $u^{(\alpha)}_{{\bf k}}({\bf r},t)$ has relations in the symmetric gauge: 
\begin{eqnarray}
u^{(\alpha)}_{\bf k} ({\bf r} + {\bf a}^{\prime}, t)&=&
\exp\left[i \frac{e}{\hbar} {\bf a}^{\prime}\cdot{\bf A}(t)\right] 
u^{(\alpha)}_{\bf k} ({\bf r}, t), 
\nonumber\\
u^{(\alpha)}_{\bf k} ({\bf r} + q {\bf b}^{\prime}, t)&=&
\exp\left[i \frac{e}{\hbar} q {\bf b}^{\prime}\cdot{\bf A}(t)\right] u^{(\alpha)}_{\bf k} ({\bf r}, t),   
\nonumber\\
u^{(\alpha)}_{\bf k} ({\bf r} + {\bf c}^{\prime}, t)&=&
\exp\left[i \frac{e}{\hbar} {\bf c}^{\prime} \cdot {\bf A}(t)\right]  u^{(\alpha)}_{\bf k} ({\bf r}, t). 
\label{Bloch-periodic-3D}
\end{eqnarray}
As well as 2D, we have a relation 
\begin{equation}
u^{(\alpha)}_{\bf k}(t)=u^{(\alpha)}_{{\bf k} - t \frac{e {\bf E} }{\hbar}}. 
\label{u(t)}
\end{equation}

According to Berry\cite{Berry}, in the adiabatic approach, 
the solution for the time-dependent Shr{\" o}dinger equation 
$i \hbar \partial \Psi(t) / \partial t = H(t) \Psi (t)$ 
can be written with a phase $\gamma^{(\alpha)}_{\bf k}(t)$ as 
\begin{eqnarray}
\Psi^{(\alpha)}_{\bf k}(t,{\bf r})
&=&\exp\left[- i \int_0^t dt^{\prime} E^{(\alpha)}_{\bf k} (t^{\prime})\right]
\times
\nonumber\\
&&\exp\left[i \gamma^{(\alpha)}_{\bf k} (t) \right] 
e^{i {\bf k}\cdot {\bf r}} u^{(\alpha)}_{\bf k}(t,{\bf r}).   
\label{wave-function}
\end{eqnarray}

We will discuss the presence of the Berry phase later, 
and first, we show the time derivative of the macroscopic polarization 
is equivalent to the quantized Hall current in 3D\cite{Kohmoto-Halperin-Wu} as follows. 
The polarization is defined by the 3D generalization of Eq. (\ref{polar}). 
The definition agrees with that 
in Ref.\cite{King-Smith-and-Vanderbilt-Resta}.
Generalizing the derivation of Eq. (\ref{dp/dt-2}) to 3D case, 
we obtain the equation 
\begin{eqnarray}
\frac{\partial {\bf P}}{\partial t}&=&
\frac{e^2}{2 \pi h} {\bf E}\times {\bf D}, 
\nonumber\\
{\bf D}&=&\frac{1}{2 \pi i} \sum_{\alpha \leq \alpha_{\rm F}} 
\int_{\rm MBZ} d^3 k [\nabla_{\bf k} \times \tilde{\bf A}^{(\alpha)}({\bf k})],  
\label{dp-dt-0}
\end{eqnarray}
where we introduce a ``gauge field'' defined as 
$\tilde{\bf A}^{(\alpha)}({\bf k})=
\left< u_{\bf k}^{(\alpha)}|{\bf \nabla}_{\bf k}|u_{\bf k}^{(\alpha)}
\right>$, which is the 3D extension of Eq. (\ref{Berry-connection-gauge}). 
What we have to do here is 
to investigate the form for the three dimensional vector ${\bf D}$. 
We expand ${\bf D}$ by the fundamental reciprocal vectors 
${\bf G}_{a^{\prime}}$,
${\bf G}_{b^{\prime}}$ and ${\bf G}_{c^{\prime}}$. 
The coefficient of the vector 
${\bf G}_{c^{\prime}}$ is written as 
\begin{eqnarray} 
t_{c^{\prime}}&=&\frac{1}{2 \pi}({\bf c}^{\prime} \cdot {\bf D})
\nonumber\\
&=& \sum_{\alpha \leq \alpha_{\rm F}} 
\int^1_0 d f_3 \int_{S(f_3)} \frac{d^2 k}{2 \pi i} 
\left[\nabla_{\bf k} \times \tilde{\bf A}^{(\alpha)}({\bf k})\right]
\cdot \frac{{\bf c}^{\prime}}{|{\bf c}^{\prime}|}.  
\label{t_c}
\end{eqnarray}
Here, we use the relation $\int_{MBZ} d^3k = \int^1_0 d f_3 ({\bf
G}_{c^{\prime}} \cdot {\bf c}^{\prime}/|{\bf c^{\prime}}|)\int_{S(f_3)} d^2 k$
, where $S(f_3)$
denotes the plane 
between two fundamental reciprocal vectors ${\bf G}_{a^{\prime}}$
and ${\bf G}_{b^{\prime}}$ (See, Eq. (\ref{crystal-momentum})). 
The r.h.s. of Eq. (\ref{t_c}) is essentially equivalent to the Chern
number\cite{Kohmoto-85,topology-text,Kohmoto-Halperin-Wu}. 
We can show in the same manner that 
$t_{a^{\prime}}={{\bf a}^{\prime} \cdot {\bf D}} /
2 \pi $ and 
$t_{b^{\prime}}=q ({\bf b}^{\prime} \cdot {\bf D}) / 2 \pi$ are also equivalent
to the Chern number. Because of the topological nature, these three numbers
$t_{a^{\prime}}$ $t_{b^{\prime}}$ and $t_{c^{\prime}}$ become integer 
whenever the Fermi energy lies in the energy Gap, 
even if the magnetic field is not rational. By choosing a different MBZ, 
in which the roles of ${\bf a}^{\prime}$ and ${\bf b}^{\prime}$ are
exchanged, we can find that ${\bf b}^{\prime} \cdot {\bf D} / 2 \pi$ 
is itself an integer. Therefore, ${\bf D}$ is a vector on 
the reciprocal 
lattice generated by ${\bf G}_{a^{\prime}}$, ${\bf G}_{b^{\prime}}$ and
${\bf G}_{c^{\prime}}$ written as 
\begin{equation} 
{\bf D}={\bf G}=-(t_{a^{\prime}} {\bf G}_{a^{\prime}}
+t_{b^{\prime}} {\bf G}_{b^{\prime}}+t_{c^{\prime}} {\bf G}_{c^{\prime}}),  
\label{D=G}
\end{equation}
i.e.
\begin{equation}
\frac{\partial {\bf P}(t)}{\partial t}=\frac{e^2}{2 \pi h} {\bf E} \times {\bf G}. 
\label{dp/dt-3}
\end{equation}
Eq. (\ref{D=G}) and (\ref{dp/dt-3}) suggest that $\partial {\bf P} /
\partial t$ is actually the quantized Hall current in
3D\cite{Kohmoto-Halperin-Wu}.

Then, we will derive the relation between the Berry
phase and the Chern number in 3D. This is the 3D generalization of 
the discussion in Ref.\cite{Kohmoto-93}. Recently, the calculation is noted by 
the authors\cite{Goryo-Kohmoto-1}.  

By substituting 
Eq. (\ref{wave-function}) into the time-dependent Schr{\"o}dinger equation 
$$ i \hbar \frac{\partial \Psi^{(\alpha)}_{\bf k}(t)}{\partial t} =H(t)
\Psi^{(\alpha)}_{\bf k}(t),$$ 
$\gamma^{\alpha}_{\bf k}(t)$ 
is obtained as 
\begin{eqnarray}  
\gamma^{(\alpha)}_{{\bf k}} (t)&=&i \int_0^t d t^{\prime} 
\left< u^{(\alpha)}_{\bf k}(t^{\prime})\left|
\frac{\partial}{\partial t^{\prime}}\right|
u^{(\alpha)}_{\bf k} (t^{\prime})\right>  
\nonumber\\
&=&i \int_0^t d t^{\prime} 
\left< u^{(\alpha)}_{{\bf k}-e {\bf E}t^{\prime} / \hbar}\left|\frac{\partial}{\partial t^{\prime}}\right|
u^{(\alpha)}_{{\bf k}-e{\bf E}t^{\prime} / \hbar}\right>.
\label{gamma_n}
\end{eqnarray}
In order to consider the Berry phase, a Hamiltonian must go around a
 closed loop in a parameter space in adiabatic process. The Hamiltonian
 $H_{\bf k}(t)$ as it is not have this property. However, it is possible 
to compactify it into the magnetic Brillouin zone 
as $H_{f_1 + 1, f_2, f_3}(t) \sim 
H_{f_1, f_2 + 1, f_3}(t) \sim H_{f_1, f_2, f_3 + 1}(t) \sim 
H_{f_1, f_2, f_3}(t)$, where 
$f_1$, $f_2$ and $f_3$ parameterizes ${\bf k}$ as 
 Eq. (\ref{crystal-momentum}) \cite{Zak}. 
Repeating the discussions in 2D system 
and by using Eq. (\ref{Berry-connection}), 
Berry phases in the case that ${\bf E}//{\bf G}_{a^{\prime}}$,
${\bf E}//{\bf G}_{b^{\prime}}$ and ${\bf E}//{\bf G}_{c^{\prime}}$,  
are written as 
\begin{eqnarray}
\Gamma^{(\alpha)}_{a^{\prime}} (f_{b^{\prime}},f_{c^{\prime}})&=&
i \oint_{C(f_{b^{\prime}},f_{c^{\prime}})} d {\bf k} 
\cdot \tilde{\bf A}^{(\alpha)}({\bf k}) 
\nonumber\\
&=&i \int_0^{1} d f_{a^{\prime}} 
{\bf G}_{a^{\prime}} \cdot \tilde{\bf A}^{(\alpha)}({\bf k}),  
\label{Berry-phase-a}\\
\Gamma^{(\alpha)}_{b^{\prime}} (f_{c^{\prime}},f_{a^{\prime}})
&=&i \oint_{C(f_{c^{\prime}},f_{a^{\prime}})} 
d {\bf k} \cdot \tilde{\bf A}^{(\alpha)}({\bf k})
\nonumber\\
&=&i \int_0^{1} \frac{d f_{b^{\prime}}}{Q} {\bf G}_{b^{\prime}} \cdot 
\tilde{\bf A}^{(\alpha)}({\bf k}), 
\label{Berry-phase-b}
\end{eqnarray}
\noindent
and 
\begin{eqnarray}
\Gamma^{(\alpha)}_{c^{\prime}}(f_{a^{\prime}},f_{b^{\prime}})
&=&i \oint_{C(f_{a^{\prime}},f_{b^{\prime}})} 
d {\bf k} \cdot \tilde{\bf A}^{(\alpha)}({\bf k})
\nonumber\\
&=&i \int_0^{1} d f_{c^{\prime}} {\bf G}_{c^{\prime}} 
\cdot \tilde{\bf A}^{(\alpha)}({\bf k}), 
\label{Berry-phase-c}
\end{eqnarray}
respectively. Here, 
$\oint_{C(f_i, f_j )}, ~(i,j=a^{\prime},b^{\prime},c^{\prime})$ denotes 
that the path of the line integral is taken on a loop 
where parameters $f_i$ and $f_j$ are fixed.    
In general, we can define the Berry phase 
when the electric field is parallel to the reciprocal lattice vector for
the magnetic unit cell,
i.e., 
\begin{equation}
{\bf E} ~//~ (l {\bf G}_a + m {\bf G}_b/ q + n {\bf G}_c),  
\label{commensulate-3D}
\end{equation}
where $l,m,n=0,\pm1,\pm2\cdot\cdot\cdot$. 

We apply the Stokes theorem in Eq. (\ref{t_c}) and one of the three
integers $t_{c^{\prime}}$, which is equivalent to the Chern number, can 
also be written in terms of Berry phase as 
\begin{eqnarray}
t_{c^{\prime}}&=&
\frac{1}{2 \pi}\sum_{\alpha \leq \alpha_{\rm F}} 
\int_0^1 d f_{c^{\prime}} 
\left[\int_0^1 d f_{a^{\prime}} \frac{d}{d f_{a^{\prime}}} 
\Gamma^{(\alpha)}_{b^{\prime}}(f_{c^{\prime}},f_{a^{\prime}}) 
- \right.
\nonumber\\
&&\left.\int_0^1 d f_{b^{\prime}} 
\frac{d}{d f_{b^{\prime}}} 
\Gamma^{(\alpha)}_{a^{\prime}} (f_{b^{\prime}}, f_{c^{\prime}}) \right], 
\label{t-c-3D}
\end{eqnarray}
and also $t_{a^{\prime}}$, $t_{b^{\prime}}$ are written as 
\begin{eqnarray}
t_{a^{\prime}}&=&\frac{1}{2 \pi}\sum_{\alpha \leq \alpha_{\rm F}} 
\int_0^1 d f_{a^{\prime}} 
\left[\int_0^1 d f_{b^{\prime}} \frac{d}{d f_{b^{\prime}}} 
\Gamma^{(\alpha)}_{c^{\prime}}(f_{a^{\prime}},f_{b^{\prime}}) 
- \right.
\nonumber\\
&&\left.\int_0^1 d f_{c^{\prime}} \frac{d}{d f_{c^{\prime}}} 
\Gamma^{(\alpha)}_{b^{\prime}} (f_{c^{\prime}}, f_{a^{\prime}}) \right], 
\label{t-a-3D}\\
t_{b^{\prime}}&=&
\frac{1}{2 \pi}\sum_{\alpha \leq \alpha_{\rm F}} 
\int_0^1 d f_{b^{\prime}} 
\left[\int_0^1 d f_{c^{\prime}} \frac{d}{d f_{c^{\prime}}} 
\Gamma^{(\alpha)}_{a^{\prime}}(f_{b^{\prime}},f_{c^{\prime}}) 
-\right. 
\nonumber\\
&&
\left.\int_0^1 d f_{a^{\prime}} \frac{d}{d f_{a^{\prime}}} 
\Gamma^{(\alpha)}_{c^{\prime}} (f_{a^{\prime}}, f_{b^{\prime}}) \right]. 
\label{t-b-3D}
\end{eqnarray}
Then, we calculate the macroscopic polarization change per a cycle. 
As well as 2D, we can define the cycle when ${\bf E}$ is parallel 
to the reciprocal lattice vector $l {\bf G}_a + m {\bf G}_b + n {\bf G}_c$. 
For example, the period is written as (See, the paragraph before Eq. (\ref{Gamma-y})),
\begin{equation}
T= \left\{  
\matrix{\frac{h}{v_0} \frac{|{\bf b}^{\prime} \times {\bf c}^{\prime}|}{e E}
~{\rm for}~{\bf E} // {\bf G}_{a^{\prime}} \perp b^{\prime}c^{\prime}-{\rm plane}, \cr
\frac{h}{q v_0} \frac{|{\bf c}^{\prime} \times {\bf a}^{\prime}|}{e E} 
~{\rm for}~{\bf E} // {\bf G}_{b^{\prime}} \perp c^{\prime}a^{\prime}-{\rm plane}, \cr 
\frac{h}{v_0} \frac{|{\bf a}^{\prime} \times {\bf b}^{\prime}|}{e E} 
~{\rm for}~{\bf E} // {\bf G}_{c^{\prime}} \perp a^{\prime}b^{\prime}-{\rm plane}.  
}
\right.
\end{equation}
Then, the polarization change per $T$ is given; 
\begin{equation}
\Delta {\bf P}= \left\{  
\matrix{\frac{e}{q v_0} (q {\bf b}^{\prime} t_{c^{\prime}}- {\bf c}^{\prime} q t_{b^{\prime}} ) 
~{\rm for}~{\bf E} \perp b^{\prime}c^{\prime}-{\rm plane},\cr
\frac{e}{q v_0} ({\bf c}^{\prime} t_{a^{\prime}}- {\bf a}^{\prime} t_{c^{\prime}} ) 
~{\rm for}~{\bf E} \perp c^{\prime}a^{\prime}-{\rm plane},\cr 
\frac{h}{q v_0} ({\bf a}^{\prime} q t_{b^{\prime}}- q {\bf b}^{\prime} t_{a^{\prime}} )   
~{\rm for}~{\bf E} \perp a^{\prime}b^{\prime}-{\rm plane}. 
}
\right.
\end{equation}
For Eqs. (\ref{t-c-3D}),(\ref{t-a-3D}) and (\ref{t-b-3D}), the polarization 
is written in terms of the Berry phase. The dipole moment 
per the magnetic unit cell i.e. $q v_0 \Delta {\bf P}$ is quantized in
each cases. 

Therefore, one can see that the polarization in 3D also has the properties \ref{a},
\ref{b}, \ref{c}, \ref{d}, and \ref{e} written in the previous section. 
Then, the polarization is quite different from the usual dielectric
polarization and has similarity with the spontaneous polarization in 
dielectric crystalline\cite{King-Smith-and-Vanderbilt-Resta}.  
As a property that it is absent in 2D, one can point out 
that the polarization changes its direction discretely in the 2D plane perpendicular 
to ${\bf E}$, when one changes the location of the Fermi level. The fact 
reflects the feature of the 3D quantized Hall effect\cite{Kohmoto-Halperin-Wu}. 
 
As well as 2D, the quantization of the dipole moment per the 
magnetic unit cell per $T$
implies the quantized charge transfer 
across the boundary of the magnetic unit cell per $T$. 
As the simple extension of Eq. (\ref{charge-2D}) to 3D, 
one can obtain the charge transfer    
\begin{equation}
\Delta Q=
\left\{
\matrix{- e (q t_{b^{\prime}} - t_{c^{\prime}} ) 
~{\rm for}~{\bf E}\perp b^{\prime}c^{\prime}-{\rm plane}, \cr 
-e (t_{c^{\prime}} - t_{a^{\prime}}) 
~{\rm for}~{\bf E}\perp c^{\prime}a^{\prime}-{\rm plane},\cr 
-e (t_{a^{\prime}} - q t_{b^{\prime}}) 
~{\rm for}~{\bf E}\perp a^{\prime}b^{\prime}-{\rm plane}.\cr 
}
\right. 
\end{equation}
We should note the fact that the charge transfer is caused by the 
dc Hall current and the results does not depend on $T$.  
Then, the results is similar to the adiabatic pumping 
with the commensurate periodic ac perturbation\cite{Thouless-pump}. 
The commensurability seems to correspond to the direction of ${\bf E}$ in our 
discussion (See. Eq. (\ref{commensulate-3D})), which is essential 
to define $T$ and $\Delta Q$. 


\section{summary}


In this paper, we have considered Bloch electrons in the presence of 
uniform electromagnetic fields in 2D and 3D. 
The Berry phase has been induced by the adiabatic change of the time-dependent vector potential.  
Following Ref.\cite{King-Smith-and-Vanderbilt-Resta}, 
the electric macroscopic polarization has been defined in the system.  
It has been shown that the time derivative of the macroscopic electric 
polarization corresponds to the quantized Hall current in the system 
whose conductivity is represented by the Chern number. 
Recently, it was shown that the Hall conductivity  
is written by the Berry phase as well as the Chern number 
in the 2D systems\cite{Kohmoto-93}, and also in 3D systems\cite{Goryo-Kohmoto-1}.
Then, we have found out that the macroscopic polarization is closely
related to the Berry phase. 
The quantized Hall current and also the polarization has been  
induced adiabatically. In the adiabatic process, the system  
always remains in the ground state, and then, the polarization in our
system is quite different from the usual dielectric polarization. 
The relation between the macroscopic polarization 
and the Berry phase is analogous to that between the spontaneous
polarization in the crystalline 
dielectrics and the Berry phase\cite{King-Smith-and-Vanderbilt-Resta}.  
We have also argued the relation between our results 
and the adiabatic pumping, which is discussed 
progressively\cite{Thouless-pump,Thouless-pump-2}.  
    
We have discussed the analogy among 
the polarization in the quantum Hall system, 
the spontaneous polarization in dielectric crystalline 
and the adiabatic pumping. 
As an essential point, the analogy comes from 
the fact that these effects are caused by the closed adiabatic change   
in the Bloch electron systems with the finite energy gap.

\acknowledgements     
The authors are grateful to H. Aoki, K. Ishikawa, N. Maeda, 
M. Sato, and F. Zhou for useful discussions.

\end{multicols}
\end{document}